\documentclass[12pt]{article}
\usepackage{graphicx}
\usepackage{epsfig}
\newcommand{\beq}{\begin{equation}}
\newcommand{\eeq}{\end{equation}}
\newcommand{\beqa}{\begin{eqnarray}}
\newcommand{\eeqa}{\end{eqnarray}}
\newcommand{\beqar}{\begin{eqnarray*}}
\newcommand{\eeqar}{\end{eqnarray*}}

\newcommand{\labell}[1]{\label{#1}} 
\newcommand{\reef}[1]{(\ref{#1})}

\newcommand{\eg}{{\it e.g.,}\ }
\newcommand{\ie}{{\it i.e.,}\ }

\newcommand{\norm}[1]{\raise.3ex\hbox{:}#1\raise.3ex\hbox{:}}

\newcommand{\gsim}{\mathrel{\raisebox{-.6ex}{$\stackrel{\textstyle>}{\sim}$}}}
\newcommand{\lsim}{\mathrel{\raisebox{-.6ex}{$\stackrel{\textstyle<}{\sim}$}}}

\newcommand{\cA}{{\cal A}}

\begin{document}

\setlength{\unitlength}{1mm}

\thispagestyle{empty}
\rightline{\small hep-th/0308056 \hfill}
\vspace*{3cm}

\begin{center}
{\bf \Large Instability of Ultra-Spinning 
Black Holes}\\
\vspace*{2cm}

{\bf Roberto Emparan}\footnote{E-mail: {\tt emparan@ffn.ub.es}}
{\bf and Robert C. Myers}\footnote{E-mail: {\tt
rmyers@perimeterinstitute.ca}}

\vspace*{0.2cm}

{\it $^1\,$Departament de F{\'\i}sica Fonamental, and}\\
{\it C.E.R. en Astrof\'{\i}sica, F\'{\i}sica de Part\'{\i}cules i Cosmologia,}\\
{\it Universitat de Barcelona, Diagonal 647, E-08028 Barcelona, Spain}\\[.5em]

{\it $^1\,$Instituci\'o Catalana de Recerca i Estudis Avan\c cats (ICREA)}\\
[.5em]

{\it $^2\,$Perimeter Institute for Theoretical Physics}\\
{\it 35 King Street North, Waterloo, Ontario N2J 2W9, Canada}\\[.5em]

{\it $^2\,$Department of Physics, University of Waterloo}\\
{\it Waterloo, Ontario N2L 3G1, Canada}\\[.5em]

\vspace{2cm} {\bf ABSTRACT} 
\end{center}

It has long been known that, in higher-dimensional general relativity,
there are black hole solutions with an arbitrarily large angular
momentum for a fixed mass \cite{MP}. We examine the geometry of the
event horizon of such ultra-spinning black holes and argue that these
solutions become unstable at large enough rotation. Hence we find that
higher-dimensional general relativity imposes an effective
``Kerr-bound'' on spinning black holes through a dynamical decay
mechanism. Our results also give indications of the existence of
new stationary 
black holes with ``rippled" horizons of
spherical topology. We consider various scenarios for the possible 
decay of ultra-spinning black holes, and finally discuss the implications
of our results for black holes in braneworld scenarios.

\vfill \setcounter{page}{0} \setcounter{footnote}{0}
\newpage

\section{Introduction}
\labell{intro}

General relativity in higher dimensions is an active area of
ongoing studies in both string theory and particle theory. In
particular, investigations over the past fifteen years have produced an
impressive catalogue of black hole solutions for various effective
theories of Einstein gravity coupled to many different kinds of matter fields
\cite{revue}. However, there have long been indications that the
physics of event horizons in higher-dimensional general relativity
is far richer and complex than in the standard four-dimensional
theory.

A simple example illustrating this point is the theorem forbidding
``topological hair'' for four-dimensional black holes. That is, each
connected component of a stationary event horizon must have the topology
of a two-sphere \cite{round}. Of course, this result is easily evaded in
higher dimensions. As an example, consider the four-dimensional
Schwarzschild metric combined with a flat metric on ${\bf R}^n$. This
space-time is an extended black ``brane'' solution of Einstein's
equations in 4+$n$ dimensions, and the topology of the horizon is
$S^2\times {\bf R}^n$. Recently, an even more dramatic violation was
presented in ref.~\cite{ER}. There, an asymptotically flat solution was
constructed describing a rotating black ring in five dimensions, with
horizon topology $S^2\times S^1$. In fact for certain values of mass and
spin, one finds that
there exist {\it three} solutions, a black hole with an
$S^3$ horizon and two different black rings with an $S^2\times S^1$ horizon. This
is a clear indication that, unlike four dimensions, in higher dimensions
black hole solutions will not be completely determined by a few
asymptotic charges (such as the mass and angular momentum).

Further, in contrast to the stability proven for four-dimensional black
holes \cite{stable4}, Gregory and Laflamme made the surprising discovery
that the extended black branes appearing in higher dimensions are
unstable \cite{GL}. Ten years later, the question of the endpoint
reached with the onset of this instability remains unresolved
\cite{chop}. However, this question lead to a conjecture that
inhomogeneous black brane solutions may also exist \cite{hm}. This 
recent conjecture
produced a surge of activity in this area \cite{bunch},
including the discovery of, at least, a certain class of such solutions
\cite{wow}. Hence, while in four dimensions black hole solutions are
extremely constrained by the uniqueness theorems, such results do not
appear to apply for higher dimensions. Rather, it seems that we
currently have only glimpses of a rich landscape of solutions with vast
unexplored areas.

In this paper, we present a preliminary study of the stability of
rotating black holes in higher dimensions.
In particular, it was discovered over
fifteen years ago that Einstein's equations in higher dimensions
have solutions describing black holes with an arbitrarily large
angular momentum for a fixed mass \cite{MP}. We refer to such
solutions as ``ultra-spinning'' black holes. No such black holes
arise in four-dimensional relativity where all solutions satisfy
the famous Kerr bound, $J\le GM^2$. However, we will argue that
the ultra-spinning black holes are in fact unstable and so an
effective Kerr bound seems to arise in higher dimensions
through a dynamical decay mechanism.

Since the analytic theory of the perturbations of
higher-dimensional rotating black holes has not been fully
developed yet, our approach is heuristic. We anticipate that our
conclusions will be useful in guiding future research in the area.
In particular, we have identified a limit (of infinite rotation)
for these black holes where an instability is known to occur, and
therefore it would make sense to focus on the subsector of the
perturbations where these unstable modes are expected.

Another consequence of our study is an indication of the
existence of new rotating black holes, of spherical topology,
where the horizon is distorted by ripples along the polar direction in
a way that preserves the same symmetries as the smoother black holes
already known.

The remainder of this paper is organized as follows: Section
\ref{thesolutions} reviews the metric for a rotating black hole in $d$
dimensions with one nonvanishing spin parameter. The focus of the
following discussion will be $d\ge6$. In section \ref{geometry}, we
examine the geometry of the event horizon of this solution when the spin
becomes large. We show that the horizon takes the shape of a 
(higher-dimensional) ``pancake'', spreading out in the plane of rotation while
becoming narrowly contracted in the other spatial directions. Section
\ref{limit} presents a limit in which this black hole geometry becomes
an unstable black membrane extending over the original plane of
rotation. We argue that the instability appearing in this limit must
also be relevant in the ultra-spinning regime. In section \ref{thermo}
we briefly comment on the thermodynamics of ultra-spinning black holes.
Section \ref{decay} considers the possible decay of the ultra-spinning
black hole by fragmentation into
multiple black holes. Finally, we conclude in section \ref{discuss} with a
discussion of the implications of our result. While throughout the main
text we focus on higher-dimensional black holes rotating in a single
plane, this is only to simplify the presentation. In the appendix \ref{multi},
we demonstrate how an analogous black brane limit can be taken for black
holes rotating in multiple planes and argue that hence instabilities
will also arise in cases when several spin parameters become large
simultaneously. Appendix \ref{radiation} presents some calculations
related to the possible decay of ultra-spinning black holes by the emission
of gravitational radiation.

\section{Ultra-spinning black holes}
\labell{thesolutions}

To begin our considerations of spinning black holes in higher
dimensions, we recall the description of angular momentum in higher
dimensions. In $d$ dimensions, the rotation group is $SO(d-1)$ for which
there are\footnote{We use the notation $\lfloor s\rfloor$ to denote the
integer part of $s$.} $\lfloor (d-1)/2\rfloor$ distinct Casimirs which
could be used to characterize the angular momentum of a system. More
concretely, the latter is described by an angular momentum two-form. By
going to the center-of-mass frame and then making a suitable rotation of
the spatial coordinates, this two-form can be put in a block-diagonal
form containing $\lfloor (d-1)/2\rfloor$ parameters $J_i$. Each of the
$J_i$ corresponds to the angular momentum associated with motion or
rotation in distinct orthogonal (spatial) planes of the 
higher-dimensional spacetime.

Hence, a black hole in $d$ dimensions is in general
characterized by $\lfloor (d-1)/2\rfloor$ $J_i$, corresponding to
spins in orthogonal planes. While in four dimensions this counting
yields the usual single $J$, for $d\ge5$ the general solution
contains two or more spins. This prolonged review simply
sets the stage for our statement that, throughout the following,
we will focus our discussion to higher-dimensional black holes
rotating in a single plane. (Further, for physical reasons that
will become apparent below, the discussion will be restricted to
$d\ge6$.) However, we stress that this restriction is made only to
simplify the presentation. In appendix \ref{multi}, we will extend the
discussion to black holes with multiple spins and in particular we
demonstrate how analogous instabilities can arise in this case.

Myers and Perry found general solutions describing asymptotically
flat black holes in $d$ dimensions with all $\lfloor
(d-1)/2\rfloor$ $J_i$ nonvanishing \cite{MP}. In the case with a
single non-zero spin $J_1=J\neq 0$ (and $J_{i>1}=0$), the metric
reduces to
%
%
\begin{eqnarray}
ds^2&=& -dt^2 + \frac{\mu}{r^{d-5}\rho^2}\left( dt+a\sin^2\theta
\,d\varphi\right)^2 +{\rho^2\over\Delta}dr^2+\rho^2d\theta^2
+(r^2+a^2)\sin^2\theta\, d\varphi^2 \nonumber\\
&&\qquad\qquad+ r^2\cos^2\theta\, d\Omega^2_{(d-4)}\,,
\labell{mphole}
\end{eqnarray}
where
\beq \rho^2=r^2+a^2\cos^2\theta\,,\qquad
\Delta=r^2+a^2-\frac{\mu}{r^{d-5}}\,. \eeq
The physical mass and angular momentum are related to the
parameters $\mu$ and $a$ by
\beq M= \frac{(d-2)\Omega_{d-2}}{16\pi G}\mu\,,\qquad
J=\frac{2}{d-2}M a\,, \labell{mandj}\eeq
where $\Omega_{d-2}$ is the area of a unit $(d-2)$-sphere. Hence
one can think of $a$ as essentially the angular momentum per unit
mass.

The first line in eq.~\reef{mphole} looks very much like the metric of
its four-dimensional counterpart, the Kerr solution, with the $1/r$
fall-off replaced, in appropriate places, by $1/r^{d-3}$ --- of course,
with the choice $d=4$ this is precisely the Kerr metric. The second line
contains the line element on a ($d$--4)-sphere which accounts for the
additional spatial dimensions. Heuristically, we can see the competition
between gravitational attraction and centrifugal repulsion by
considering the expression
\beq
\frac{\Delta}{r^2}-1=-\frac{\mu}{r^{d-3}}+\frac{a^2}{r^2}\,. \eeq
Roughly, the first term on the right-hand side corresponds to the
attractive gravitational potential and falls off in a
dimension-dependent fashion. In contrast, the repulsive
centrifugal barrier described by the second term does not depend
on the total number of dimensions, since rotations always refer to
motions in a plane. As a consequence, the features of the event
horizons will be strongly dimension-dependent.

The outer event horizon can be determined as the largest (real)
root $r_+$ of $g_{rr}^{-1}=0$ or simply $\Delta(r)=0$. That is,
\beq r_+^2+a^2 -\frac{\mu}{r_+^{d-5}}=0\,.\labell{horz}\eeq
For $d=4$, a regular horizon is present for values of the spin parameter
$a$ up to the Kerr bound: $a=\mu/2$ (or $a=GM$), which corresponds to an
extremal black hole with a single degenerate horizon (with vanishing
surface gravity). Solutions with $a>GM$ correspond to naked
singularities. In $d=5$, the situation is apparently quite similar since
the real root at $r_+=\sqrt{\mu-a^2}$ exists only up to the extremal
limit $\mu=a^2$. However, this extremal solution has zero area, and in
fact, it is a naked ring singularity. Note that the five-dimensional
black ring of ref.\ \cite{ER} evades this bound. We return to a brief
discussion of five dimensions in section \ref{discuss}.

For $d\geq 6$, $\Delta(r)$ is always positive at large values of
$r$, but the term $-\mu/r^{d-5}$ makes it negative at small $r$.
Therefore $\Delta$ always has a (single) positive real root
independent of the value of $a$. Hence regular black hole
solutions exist with arbitrarily large $a$, \ie arbitrarily large
angular momentum per unit mass. We refer to such solutions with
large angular momentum per unit mass as ``ultra-spinning'' black
holes. In the general situation with multiple spin parameters, an
ultra-spinning black hole would be one for which any of $a_i$
satisfies $a_i^{d-3}>\mu$. As will be discussed in appendix \ref{multi},
it is certainly possible to have a regular horizon where several
of the spin parameters satisfy this inequality.

\section{Geometry of ultra-spinning horizons}
\labell{geometry}

Now we consider the shape of the event horizon of an
ultra-spinning black hole. To do so, we compare the sizes of
invariant geometric quantities such as the proper areas of
different sections of the horizon.

In the limit of very large $a$ and fixed mass, the coordinate
radius of the horizon is well approximated by
\beq r_+\simeq
\left(\frac{\mu}{a^2}\right)^{1/(d-5)}\ll a\,. \labell{rplus}\eeq
Observe that $r_+$ shrinks as $a$ is increased keeping the mass
fixed. However, $r_+$ by itself does not have any invariant
meaning.

The total, $(d-2)$-dimensional area of the horizon is
\beq
\cA=\Omega_{d-2} r_+^{d-4}(r_+^2+a^2) \simeq \Omega_{d-2}
r_+^{d-4}a^2 \simeq \Omega_{d-2}\left(
\frac{\mu^{d-4}}{a^2}\right)^{1/(d-5)}\,.
\labell{atot}\eeq
Note that this area {\it decreases} as angular momentum increases
with a fixed mass. This is also the case in $d=4$, but now for higher $d$
we can consider the limit $a\to\infty$ and we see that the
area shrinks to zero in the limit.

Now consider the area of a two-dimensional section of the horizon
obtained by considering the solution \reef{mphole} at a fixed
point in the ``transverse" sphere $\Omega_{d-4}$. This
``parallel" two-dimensional area is
\beq \cA_{\parallel}^{(2)}=\Omega_2 (r_+^2+a^2)\simeq \Omega_2
a^2\,. \labell{apar}\eeq
This area {\it grows} for ultra-spinning black holes.
Alternatively, one might fix the angles $\theta$ and $\varphi$ to
consider the area of the transverse sphere with the result
\beq \cA^{(d-4)}_\perp=\Omega_{d-4}(r_+\cos\theta)^{d-4}\,.
\labell{aperp}\eeq

These observations are easily compatible if we consider that the
horizon has a characteristic size $\ell_\parallel$ in directions
parallel to the rotation plane, and $\ell_\perp$ in directions
perpendicular to it. Then, if $\ell_\parallel\gg \ell_\perp$, we
expect that
\beqa
\cA&\propto& \ell_\parallel^2\, \ell_\perp^{d-4}+\dots\labell{atot2}\\
\cA_{\parallel}^{(2)}&\propto &
\ell_\parallel^2+\dots\nonumber\\
\cA^{(d-4)}_\perp&\propto&\ell_\perp^{d-4}+\ldots\,.
\nonumber\eeqa
where the dots denote terms small in an expansion in
$\ell_\perp/\ell_\parallel$. Comparing with eqs.~\reef{atot} to
\reef{aperp} above, we easily identify
\beqa
\ell_\parallel &\sim& a\,,\nonumber\\
\ell_\perp &\sim& r_+\,.
\labell{sizes}\eeqa
%
This result
is further supported by examining the capture of null geodesics
in the plane of rotation --- see appendix \ref{zeroemission}. So
ultra-spinning black holes are characterized by very large
$\ell_\parallel$ and very small $\ell_\perp$: the horizon is
highly ``pancaked'' along the plane of rotation. That is, in
accord with the naive intuition, the horizon of these rapidly
rotating black holes spreads out in the plane of rotation while
contracting in the transverse directions.

\section{Black membrane limit}
\label{limit}

We now introduce a limit where the rotation parameter $a$ grows to
infinity for the metric \reef{mphole} with $d\ge6$. If we keep the
mass (\ie $\mu$) finite as $a\to\infty$, then the horizon radius
\reef{rplus} and area \reef{atot} shrink to zero. Since we are
interested in keeping a non-vanishing horizon, we will let $\mu\to\infty$
with $\hat \mu\equiv \mu/a^2$ held fixed.

In this limit, if we are near the horizon away from the pole
$\theta=0$, the rotation grows without bound. So in order to
produce a finite metric, we will confine ourselves to
progressively smaller regions near $\theta=0$. It is convenient to
introduce a new coordinate $\sigma= a\sin\theta$ which is kept
finite as $a\to\infty$ and therefore focuses on $\theta\to 0$. We
also keep $r$ finite. Taking this limit for the solution \reef{mphole}
we obtain
\beq ds^2=-\left(
1-\frac{\hat\mu}{r^{d-5}}\right)dt^2+\frac{dr^2}{1-\frac{\hat\mu}{r^{d-
5}}}+r^2d\Omega^2_{(d-4)}+d\sigma^2+\sigma^2 d\varphi^2\,.
\labell{memlimit}\eeq
This is the metric of a black membrane in $d$ dimensions, extended
along the plane $(\sigma,\varphi)$. The tension of the membrane is
proportional to $\hat\mu=\mu/a^2$ and so is given by the limiting
mass per unit area of the original black hole. The horizon in
eq.~\reef{mphole} began with the topology of a sphere $S^{d-2}$
and has now pancaked out in the plane of rotation producing the
topology ${\bf R^2}\times S^{d-4}$ in the limiting metric
\reef{memlimit}. The angular momentum is not longer visible in
this limit since
\beq
g_{t\varphi}\to \frac{\hat\mu \sigma
}{r^{d-5}}\,\frac{\sigma}{a} \to 0\,,
\eeq
\ie in this limit with
$\sigma\ll a$, the rotation of the horizon becomes negligible.

The existence of this limit has a remarkable consequence. Gregory
and Laflamme (GL) have shown that black branes are classically
unstable \cite{GL}. Therefore, in the limit of infinite rotation
six- or higher-dimensional black holes go over to a unstable
configuration. It is then natural to conjecture that the
instability sets in already at finite values of the rotation, so
ultra-spinning black holes become classically unstable for large
enough $a$.

Let us now argue in more detail that the instability must be present at
large but finite rotation, and is not simply an artifact of the
the limit taken above. When $a$ is large but finite, the geometry
near the horizon and near the axis of the ultra-spinning black
hole is well approximated by the metric \reef{memlimit} for
$r,\sigma < a$. More precisely, one finds that the diagonal
components of the metric are as given above in eq.~\reef{memlimit}
with corrections which are suppressed by factors of $r^2/a^2$ or
$\sigma^2/a^2$. Similarly, as indicated above, the only
off-diagonal component $g_{t\varphi}$ is suppressed by $\sigma/a$.
Hence the spacetime geometry near the center of the ``pancake"
horizon is essentially that of the black membrane \reef{memlimit}.

The wavelength of the threshold mode of the GL instability is
$\lambda_{GL}\sim \hat\mu^{1/(d-5)}\simeq r_+$. Since in the
ultra-spinning regime $r_+\ll a$, or $\mu \ll a^{d-3}$, there
is ample room on the horizon to accommodate unstable fluctuations.
In particular, one can construct initial data for localized wave
packets containing unstable modes with support in a finite region
near the horizon of the black membrane \reef{memlimit}. It would
then be possible to import this initial data with only minor
perturbations to the pole region of the horizon of an
ultra-spinning black hole, and in this setting the future
evolution should display the same growth of unstable modes as for
the full black membrane. In any event, it seems unlikely that the
boundary conditions at the edge of the ultra-spinning horizon
could eliminate the unstable modes, since the edge can be made to
lie arbitrarily far away from the pole.

In appendix \ref{multi}, we extend the above discussion to black holes
with several rotation parameters turned on. If at least one (in
even $d$) or two (in odd $d$) of the spins are much smaller than
the rest, then these black holes can have an ultra-spinning regime
where the horizon is flattened along each of the ultra-spinning
rotation planes. For example, if $n$ spins grow to infinity (with
$n<\lfloor(d-3)/2\rfloor$), one obtains, with the appropriate
limit, a black $2n$-brane solution.

\section{Remarks on thermodynamics}
\label{thermo}

It has long been supposed that the classical GL instability has a
connection with the thermodynamic properties of the black branes. In
their original work \cite{GL}, GL gave 
an argument based on global thermodynamic stability 
to suggest that the unstable black string would fragment into a chain of
localized black holes. While this scenario is controversial \cite{hm},
sharper conjectures as the relation between the classical and local
thermodynamic stability of black branes have been recently formulated
\cite{GM}. In these discussions, the appearance of a negative specific
heat of a black brane is related to the onset of a classical
instability. Previous investigations have focused on charged black
branes \cite{charge} and little consideration has been given to spin
\cite{GWK}. In this direction, let us comment that for the Kerr black
hole, the phase transition signaled by the divergence of the specific
heat (at constant $J$) \cite{davies} is not associated to any classical
instability of the Kerr black hole, which is believed to be stable.
Instead, following \cite{GM}, one would associate this phase transition
with a transition towards classical stabilization of the Kerr black
string at a finite spin. Further work in this direction is in progress.

Before leaving this topic, however, we observe that the
thermodynamics of the spinning black holes \reef{mphole} show a
qualitative change in behavior. That is, these black holes make a
transition from behaving similar to the Kerr black hole, to
behaving like a black membrane. The simplest and perhaps clearest
quantity to consider is the black hole temperature
\beq
T=\frac{1}{4\pi}\left(\frac{2 r_+^{d-4}}{\mu}+\frac{d-5}{r_+}\right)\,.
\labell{temp}\eeq
For fixed mass and increasing spin, $r_+$ always
decreases. So, beginning from zero spin, the first term in the
r.h.s.\ of \reef{temp} dominates and $T$ always decreases, like in
the familiar case of the Kerr black hole. In $d=4$ and $d=5$ the
temperature shrinks to zero at extremality, but in $d\geq 6$ there
is no extremal limit. Instead, $T$ reaches a minimum and then
starts growing like $\sim r_+^{-1}$, as expected for the black
membrane. The minimum, where the behavior changes, can be
determined exactly as
\beq \frac{a}{r_+} =
\sqrt{\frac{d-3}{d-5}}\,. \eeq
For the dimensionless quantity $a^{d-3}/\mu$ one
finds the critical values
\beq \frac{a^{d-3}}{\mu}= 1.29\quad
(d=6); \quad 1.33\quad (d=7); \quad 1.34\quad (d=8)\,. \eeq
This analysis seems to indicate that the membrane-like behavior,
and hence the instability, occurs at a relatively low value of the
angular momentum, \ie not too far into the ultra-spinning regime.
In the next section we present further evidence for this.

\section{Death by fragmentation} 
\label{decay}

The arguments of the previous sections provide a clear indication
that ultra-spinning black holes become classically unstable as the
rotation parameter grows.
A precise determination of the angular momentum per unit mass
at which the
instability appears requires a linearized analysis of the perturbations
of the Myers-Perry black holes (which may show unstable modes other than
the ones suggested by the limiting GL instability). This is a difficult
but important task that we will not pursue in this paper, where we
remain at a more heuristic level.

Furthermore, the linearized perturbation analysis does not reveal by
itself what is the final fate of the black hole. Indeed, the fate of
black branes undergoing the GL instability has been a matter of debate
over the last few years. As we mentioned in the previous section, GL
gave a thermodynamic argument for the fragmentation of a
black string \cite{GL} and we return to this reasoning here. The
essential idea was that a black string compactified on a large enough
circle could increase the horizon area if it pinched off to become
a black hole (of the same mass) localized in the circle. Hence such
a process would be driven by the second law. Further, this thermodynamic
transition happens when the
compact circle can also fit the wavelength of GL-unstable modes,
suggesting a connection between both instabilities. So \cite{GL}
proposed that the rippling of the horizon observed at linearized order
grows until the horizon pinches off down to Planckian-size necks. At
this point, quantum gravity effects should split off the black string
into black holes. After ref.\ \cite{hm}, this picture has become more
controversial. Nevertheless, it is remarkable that the global 
thermodynamical argument appears to be linked quite precisely to the
classical instability. Certainly we are not aware of any counterexamples
to this connection. Therefore in the following we will consider the
instability of the ultra-spinning higher-dimensional black holes from
the simple point of view of the entropy arguments presented in
ref.~\cite{GL}. That is, we wish to compare the horizon area of an
ultra-spinning black hole to that of a plausible final state. Whenever
the total final horizon area is larger than the area of the initial
black hole, we take it as an indication of an instability.

We examine the possibility that the ultra-spinning black
hole breaks apart into two identical black holes carrying away the spin
as orbital angular momentum. This could happen if, \eg the horizon
develops a `bar-mode' instability which eventually disrupts the
horizon. Our approximation is rather crude since we neglect the
emission of gravitational radiation along the process, but this is
the same as in the original GL argument.  We also take the final black
holes to be non-spinning, since this maximizes the final entropy. For
the same reason, we do not describe the fragmentation into more than two
black holes. We take the final black holes to be of equal mass; it is
easy to allow for different masses, but it leads to essentially the same
results.

The initial state is characterized by the black hole mass $M$ and spin
$J$, or the parameters $\mu$ and $a$ in \reef{mandj}.
Recall that the initial area \reef{atot} is 
\beq
{\cA}_0=\Omega_{d-2}r_+^{d-4}(r_+^2+a^2)\,.
\eeq

The final state consists of two black holes, each of
mass $m$, infinitely far from each
other, and moving in antiparallel directions, with impact parameter
$2R$. The latter will be fixed later on.
If the total energy and angular momentum are $M$ and $J$, then, in the
center-of-momentum frame, the momenta of the final black holes are $\pm J/2R$, so
\beq
M=2\sqrt{m^2+{J^2\over 4 R^2}}\,.
\labell{energy}
\eeq
One can solve for $m$ as
\beq
m=\frac{1}{2}\sqrt{M^2-\frac{J^2}{R^2}}\,.
\labell{finalm}\eeq
Defining
\beq
\mu_1=\frac{16\pi G}{(d-2)\Omega_{d-2}} m
\eeq
then \reef{finalm} becomes
\beq
\mu_1=\frac{r_+^{d-5}(r_+^2+a^2)}{2}\sqrt{1-\frac{4}{(d-2)^2}\frac{a^2}{R^2}}\,.
\eeq

The final black holes are assumed to be non-rotating. They are moving
apart at asymptotically constant velocity, but the black hole area is
invariant under
boosts \cite{HMt}. So the total final area is
\beq
\cA_1=2\Omega_{d-2}\mu_1^{d-2\over d-3}.
\eeq
Our criterion for an instability
\beq
{\cA_1}/{\cA_0}>1\label{areacrit}
\eeq
becomes
\beq
\frac{1+(a/r_+)^2}{2}\left(1-\frac{4}{(d-2)^2}\frac{a^2}{R^2}\right)^{\frac{d-
2}{2}}>1\,.
\labell{master}
\eeq

It is clear that $\cA_1$ will be maximized by setting $R$ as large as
possible, i.e, when the escaping black holes carry away the angular
momentum with minimal kinetic energy and hence maximal rest mass
and maximal horizon area.
However, in the absence of further knowledge about the
fragmentation process, it would not be reasonable to assume that $R$ is
much larger than the radius of the initial horizon in the rotation
plane. For large rotation, we know that this radius is approximately
equal to $a$. From \reef{master} we see that if $R$ scales as $a$ then
the inequality will always be satisfied for large enough $a$. In fact,
this will be the case whenever $R$ grows faster than $\frac{2}{d-2} a$.
So in the ultra-spinning regime the entropic arguments seem to allow for
a rather wide margin for fragmentation into two black holes.

To obtain an estimate of the transition point, we
make a specific choice requiring
that $R$ is not larger than the horizon radius,
\beq
R\leq
\sqrt{r_+^2+a^2}\,.
\labell{maxR}\eeq
Saturating this inequality yields the following values for
eq.~\reef{master} to be satisfied:
\beq \frac{a}{r_+}\geq
1.36\quad (d=6),\quad
1.26\quad (d=7),\quad
1.20\quad (d=8),
\eeq
or equivalently,
\beq
\frac{a^{d-3}}{\mu}\geq
0.88\quad (d=6),\quad
0.97\quad (d=7),\quad
1.02\quad (d=8)\,.
\eeq
Alternative choices such as $R=k\sqrt{r_+^2+a^2}$, or $R= ka$, with $k\leq 1$,
lead to minor differences in the critical value of $a/r_+$ as long
as $k>2/(d-2)$ \ie $R$ scales faster than $\frac{2}{d-2} a$. The
quantitative differences for different values of $d$ are also small.
Obviously, these values should not be taken too literally. The important
result is that the area criterion \reef{areacrit} seems to signal an instability 
for values of $a/r_+$ not much larger than one.

We have also considered a variety of other possible final
configurations, \eg final state black holes with different
masses, or two black holes flying apart, with a third one
(rotating or not) remaining at the initial position. In all cases
the qualitative results remain as above, with the critical value
of $a$ being always a few times $r_+$.

Of course, an alternative conjecture as to the evolution of the
instability is that the ultra-spinning black holes decay by the emission
of gravitational radiation. We comment on this scenario below in the next
section and in appendix \ref{radiation}.

\section{Discussion}
\label{discuss}

We have shown that in the ultra-spinning regime of a black hole, a wide
portion of its horizon is well approximated by a flat black
membrane. Taken together with the GL instability of black branes, we
come to the conclusion that ultra-spinning black holes must be
classically unstable. We have also shown that thermodynamic arguments,
based on the second law, seem to support this view. Thus we find a
dynamical ``Kerr bound" on the spin in $d$ dimensions of the form
\beq
J^{d-3}\leq \beta_d\; G M^{d-2}\,,
\labell{newbound}\eeq
where the numerical factor
\beq
\beta_d=\frac{2^{d+1}\pi}{(d-2)^{d-2}\Omega_{d-2}}\left(\frac{a^{d-
3}}{\mu}\right)_{\rm crit}
\labell{fudge}\eeq
is fixed in $d\geq 6$ by the onset of an instability at critical values
of the parameters. The simple estimates of Secs.\ \ref{limit} and
\ref{decay} suggest that $(a^{d-3}/\mu)_{\rm crit}$ is not much larger
than a few, with a weak dependence on $d$. Note that this implies
that the numerical value of $\beta_d$ becomes small rapidly with
increasing $d$ (recall $\beta_4=1$).

Throughout the preceding, we have only considered higher-dimensional
black holes rotating in a single plane. However, we stress again that
this restriction was made only to simplify the presentation. Appendix
\ref{multi} presents an extension of the black brane limit in section
\ref{limit} to black holes with multiple spins. Hence we argue that
analogous instabilities arise for this case.

In section \ref{decay}, evidence suggesting that $(a^{d-3}/\mu)_{\rm
crit}=O(1)$ came from extending the thermodynamic arguments of
ref.~\cite{GL} and considering the possibility that an ultra-spinning
black hole could fragment into several black holes carrying orbital
angular momentum. A more conventional suggestion (which does
not rely on quantum gravity effects) would be that the rotating black
hole horizon becomes distorted, and since the black hole is rotating
gravitational waves are emitted. This radiation could by itself
cause a spin-down returning the black hole to the stable regime.
However, a detailed description of such a spin-down (or the
gravitational radiation produced in any decay process) is beyond the
scope of this paper. Appendix \ref{radiation} presents some calculations
which take some tentative steps in this direction by considering
processes in which the black hole is only slightly perturbed away from
the Myers-Perry solutions \reef{mphole}. However, it seems that these
putative decay processes are in conflict with the area theorem --- they
do not seem to produce radiation and an increase in the total area at
the same time, so they are probably forbidden. The most likely outcome
is that the instability must produce highly nonlinear distortions of the
horizon in order for gravitational radiation to provide an effective
mechanism to dissipate the angular momentum of ultra-spinning black
holes.

An important application of the limiting membrane geometry is the
identification of the sector of the
perturbations of the spinning black hole where the instability is
expected. For the limiting membrane eq.~\reef{memlimit}, GL found an
unstable tensor perturbation, regular on the horizon, and preserving the
$SO(d-3)$ symmetry of the transverse sphere $S^{d-4}$, with components of
the generic form
\beq
h^{\mu\nu}\sim e^{\Omega t} e^{i(k_1 z^1+k_2 z^2)} h(r)\,,
\labell{GLmode}\eeq
where $z^1$, $z^2$ are Cartesian coordinates along the planar membrane
direction. They found that when $k=\sqrt{k_1^2+k_2^2}$ is below a
critical wavenumber $k_c$ $(\sim r_+^{-1})$, then $\Omega$ becomes real
and positive, hence
the perturbation is unstable. At the threshold $k=k_c$, there is a static,
zero-mode perturbation with $\Omega=0$. Note that it is only $k$, and
not $k_1$ or
$k_2$ separately, that enters the perturbation. Therefore, if instead of
the plane wave basis in $(z^1,z^2)$ in \reef{GLmode}, we choose a
cylindrical basis in polar coordinates
$(\sigma,\varphi)$ with Bessel wavefunctions,
\beq
h^{\mu\nu}\sim e^{\Omega t} J_m(k \sigma) e^{im\varphi} h(r)\,,
\labell{GLmode2}
\eeq
then the azimuthal number $m$ is irrelevant for the GL perturbations, in
the sense that whenever $k<k_c$, the instability will be present for any
value of $m$. So, in particular, there do exist unstable modes in the
axially symmetric sector $m=0$. The radial profile of one such unstable
mode is a cylindrical wave $J_0(k \sigma)$.

Importing these observations
to the perturbation problem of ultra-spinning black holes,
where we trade $\sigma$ for the polar angle $\theta$, we see that in
order to identify an unstable mode we can restrict ourselves to
perturbations that depend only on $r$ and $\theta$, and which preserve
all the rotation symmetries $SO(2)\times SO(d-3)$ of the black hole. So
the instability need not break the axial symmetry of the rotating black
hole, and this unstable mode will not be radiated away, at least at this
linearized order. Other modes which are not axially symmetric presumably
become more complicated as one reaches the region of the horizon where
rotation ceases to be negligible. Such perturbations should
cause the black hole to emit gravitational waves, as considered above.

An interesting recent study of the stability problem focused on scalar
perturbations of the five-dimensional rotating black hole, and also on
axially symmetric scalar perturbations in higher dimensions\footnote{The
same equations for scalar perturbations have been obtained by a number
of other researchers \cite{miss2}. The separability of the variables $r$
and $\theta$ has been noted by all of them, but only ref.~\cite{miss}
have investigated stability in detail.}. No signs of instability were
found \cite{miss}. This is indeed compatible with our analysis, since
black branes are known to be stable to scalar perturbations \cite{GL}.
Therefore the sector where the unstable mode indicated by the limiting
GL instability is expected to lie, has not been probed yet. There is of
course the possibility of unstable perturbations besides the ones we
have identified.

Consider now the threshold, zero-mode axisymmetric perturbation with
$k=k_c$. It has been shown in \cite{wow} that the GL zero-mode signals a
new branch of black brane solutions, which break the translation
invariance along the brane. These are inhomogeneous black branes, with
an energy higher than the corresponding homogeneous branes. Their
existence suggests the possibility of a similar new class of stationary
rotating black holes with a rippled horizon of spherical topology. For
one such ultra-spinning rippled black hole, the geometry near the pole
$\theta \ll 1$ and $r\ll a$ should be well approximated by an axially
symmetric inhomogeneous black membrane. The ripples have a profile in
the polar angle $\theta$, and preserve all the rotational $SO(2)\times
SO(d-3)$ symmetry of the horizon. If one were able to develop at least
the relevant part of the perturbation theory of spinning black holes,
then the rippled rotating black holes should branch-off from
axisymmetric zero-mode perturbations. Of course, these black holes may
be unstable themselves, \eg to perturbations that break axial symmetry.

It has been conjectured that there should exist spinning black hole
solutions with only two Killing isometries \cite{harv}. However, the new
solutions we are proposing are not of this sort, since they will have
the same number of Killing isometries as the already known black holes.
Even if we are conjecturing their existence from membrane solutions that
break some symmetry, the radial symmetry that is broken is absent at any
finite values of the rotation.

If such solutions actually exist, then it would seem impossible to have
a version of black hole uniqueness within the class of black holes with
a given topology of the horizon. Our study might, perhaps, be taken to
add to the conjecture that local stability, which plays no role in the proof
of uniqueness in four dimensions, may be the feature that selects, in
higher dimensions, one solution among others with the same asymptotic
charges \cite{kol}. However, the catalogue of solutions and their
properties are too poorly understood at present to extract any conclusions in
this respect.

We have focused on $d\geq 6$, but the situation in $d=5$ is also quite
interesting.
Although these black holes cannot be ultra-spinning, their geometry
close to extremality is also a thin pancake of radius $\ell_\parallel\sim
a$ and thickness $\ell_\perp \sim r_+$. However,
there is no black brane limit, since in $d=5$ there are no black
two-branes. So there is no robust argument for an instability in this
case.

Nevertheless, it has been conjectured that an instability sets in before
the extremal solution is reached \cite{ER}.
We observe further that the argument for fragmentation in Sec.\
\ref{decay} applies as well in $d=5$. 
The extremal limit in $d=5$ corresponds to $a^2\to \mu$, $r_+\to 0$, \ie
$a/r_+\to\infty$. With the choice \reef{maxR}, the area criterion
\reef{master} implies the critical value $a/r_+= 1.60$, or
\beq
\left(\frac{a^2}{\mu}\right)_{\rm crit}= 0.72 \quad (d=5)\,.
\eeq
So even if this case has no black brane limit, an instability near
extremality could be expected. Instead, in $d=4$ we cannot have $a>r_+$
and, consistently, none of the previous arguments predict any
instability.

On the other hand, the
five-dimensional black ring of \cite{ER} does have an ultra-spinning
regime. If the limit of infinite spin is taken in such a way that
the horizon area per unit length of the ring remains finite, then it
approaches a straight, boosted black string. This is also expected to
suffer from the GL instability, and therefore a Kerr-bound similar to
\reef{newbound} should hold. Note that all known
solutions that admit an ultra-spinning regime have some form of black
brane limit at infinite rotation. This may be a generic feature.

Finally we would like to conclude with a few comments of the
implications of our results for black holes in braneworld scenarios
\cite{braneworld}. In these higher-dimensional models, the fundamental
scale of gravity is dramatically reduced, possibly even to be $O({\rm TeV})$,
and the size of the compact dimensions is increased to be much larger
than the corresponding fundamental length scale. Hence one must
consider seriously classical solutions of general relativity in higher
dimensions. One of the consequences is that there are discernible
changes in the properties of black holes when their Schwarzschild
radius is less than the compactification scale \cite{newer}. 
In section \ref{geometry},
we showed that the horizon of an ultra-spinning black hole was very
extended in the plane of rotation but very narrow in all of the transverse
directions. Hence, naively, it would seem that such black holes could
arise in braneworld models, and as a consequence, we might expect
to find {\it macroscopic} black holes whose angular momentum far exceeds
the standard Kerr bound. An observation of such a black hole would
seem to provide clear evidence of extra dimensions. Unfortunately,
this line of reasoning is undermined by our discovery that ultra-spinning
black holes are unstable. That is, even if such a macroscopic black hole
was created, it would rapidly decay and presumably settle down to
a configuration which presents no violations of the Kerr bound.
 
Note, however, that given the inequality \reef{newbound}, microscopic
black holes need not conform to the four-dimensional Kerr bound. For
comparison purposes consider six spacetime dimensions. Then
eq.~\reef{newbound} yields $J\lsim(M/M_{fun})^{4/3}$ while the standard
Kerr bound becomes $J\lsim(M/M_{Planck})^2\simeq
M^2/(M_{fun}^4L_{compact}^2)$. Hence the maximum angular momentum rises
more rapidly in the first (higher-dimensional) inequality because of
both the power and the prefactor. Of course, the expressions on the
l.h.s. of the two inequalities meet roughly when the horizon radius of
the black hole reaches the compactification scale $L_{compact}$. Hence
the microscopic or higher-dimensional Kerr bound merges smoothly
with the standard four-dimensional Kerr bound which still constrains the
macroscopic black holes.

Another dramatic prediction of the braneworld scenarios is that it
becomes much easier to produce microscopic black holes in subatomic
collisions of elementary particles --- see, for example,
refs.~\cite{bang1,bang2,spinbang}. In the initial analyses \cite{bang1},
the cross-section for black hole formation in a grazing collision was
estimated to be $\pi\,R_{S}^2$, where $R_S$ was the (higher-dimensional)
Schwarzschild radius associated with the center-of-mass energy. In many
cases \cite{spinbang} it was suggested that an improvement which
took into account the angular momentum associated with particles
colliding with a finite impact parameter was to replace $R_S$ with the
coordinate radius $r_+$ determined by eq.~\reef{horz}. However, $r_+$
has no invariant meaning as it can be changed by a coordinate
transformation, and moreover, given
the discussion of the horizon geometry in section \ref{geometry}, this
choice would seem to be a gross underestimate which does not conform
with the geometric picture that motivated the original estimate. 
More convincingly, one can employ analytic and numerical
methods \cite{apparent,app} to search for an apparent horizon for two
particles colliding with a finite impact parameter. This analysis puts
an upper bound on the impact parameter in higher dimensions 
\cite{app}.\footnote{These results have
recently been incorporated in the phenomenological analysis 
of ref.~\cite{app2}.}

Hence it appears that the ultra-spinning black holes play no role in the
black hole production from particle collisions, and it seems plausible that
this result is related to the instability of these solutions. One's
usual intuition (as developed in four dimensions) is that black holes
are stable objects and hence gravitational collapse inevitably settles
down to one of a limited family of black hole solutions. Now roughly
one may think of the solutions of Einstein's equations as trajectories
in a superspace of spatial geometries, and then the previous
intuition could be summarized by saying that the black hole
solutions are distinguished as attractors in this space of trajectories.
In contrast, the instability of the ultra-spinning black holes suggests
that these solutions actually are ``repellors" in the space of
solutions. That is, nearby trajectories are driven away by the unstable 
nature of these ultra-spinning solutions. Turning this around
then, one might interpret the results on apparent horizons \cite{app} as
further evidence that $(a^{d-3}/\mu)_{\rm crit}=O(1)$.

In any event, because of their instability, it seems that ultra-spinning
black holes have no dramatic consequences for braneworld
scenarios.


\section*{Acknowledgements}
RCM was supported in part by NSERC of Canada and Fonds FCAR du Qu\'ebec.
RE was supported in part by grants UPV00172.310-14497, MCyT
FPA2001-3598, DURSI 2001-SGR-00188 and HPRN-CT-2000-00131. RE would like
to thank the Perimeter Institute for their warm hospitality during a
portion of this work. RCM thanks the KITP for their hospitality while
this paper was being finished. Research at the KITP was supported in part by the
National Science Foundation under Grant No.~PHY99-07949.
We also thank Jordan Hovdebo and Don Marolf
for useful conversations. We have presented this work at CIAR Cosmology
and Gravity Program Meeting (February 2001); IFT-UAM, Madrid (April 2001);
CERN (May 2001); GR-16 Conference, Durban (July 2001); Feza G\"ursey
Institute, Istanbul (July 2003), and we would like to thank the
audiences at all these places for their positive response and feedback.

\appendix

\section{Black brane limit with several independent spins}
\label{multi}

We discuss first the case of an odd number of spacetime dimensions
$d$. The solution for a black hole with arbitrary rotation in each
of the $(d-1)/2$ independent rotation planes is \cite{MP}
\beq
ds^2=-dt^2+(r^2+a_i^2)(d\mu_i^2+\mu_i^2 d\varphi_i^2) +\frac{\mu
r^2}{\Pi F}(dt+a_i\mu_i^2 d\varphi_i)^2+\frac{\Pi F}{\Pi-\mu
r^2}dr^2\,. \labell{odd}\eeq
Unless otherwise stated, we assume
summation over $i=1,\dots, \frac{d-1}{2}$. The mass parameter is
$\mu$, not to be confused with the direction cosines $\mu_i$,
which satisfy $\mu_i^2=1$. Here
\beq
F=1-\frac{a_i^2\mu_i^2}{r^2+a_i^2}\,,\qquad
\Pi=\prod_{i=1}^{(d-1)/2}(r^2+a_i^2)\,. \labell{fpi}\eeq
We assume that the parameters are such that a horizon exists. A
sufficient, but not necessary, condition is that any two of the
spin parameters vanish, \ie if two $a_i$ vanish, a horizon will
always exist irrespective of how large the remaining spin
parameters are \cite{MP}.

Take the first $n$ rotation parameters to be comparable among
themselves, and much larger than the remaining $(d-1)/2-n$ ones. We
use the index $j$ for the former, and $k$ for the latter. We take the
limit
\beqa
a_j&\to \infty\,,\qquad &j=1,\dots,n\,,\nonumber\\
a_k&\  {\rm finite}\,,\qquad &k=n+1,\dots,\frac{d-1}{2}\,,
\eeqa
where the different $a_j$ go to infinity at the same rate.
We also take $\mu\to\infty$, keeping
\beq \frac{\mu}{\prod_j
a_j^2}\equiv \hat \mu \labell{hatmu}\eeq
finite, and define new coordinates $\sigma_j=a_j \mu_j$ (no sum)
that stay finite as we approach $\mu_j\to 0$. The remaining
$\mu_k$ must satisfy $\mu_k^2=1$. In the limit
\beqa
F&\to&1-\frac{a_k^2\mu_k^2}{r^2+a_k^2}\equiv {\widehat F}\,,\nonumber\\
\Pi&\to& \prod_{k}(r^2+a_k^2) \prod_j a_j^2\equiv  {\widehat \Pi}
\prod_j a_j^2\,, \eeqa
\ie ${\widehat F}$ and ${\widehat \Pi}$ are the same functions as in
\reef{fpi} but now involving only the $(d-1)/2-n$ rotation
parameters that remain finite. The metric that results is
\beqa ds^2&=&-dt^2+(r^2+a_k^2)(d\mu_k^2+\mu_k^2 d\varphi_k^2)
+\frac{\hat\mu r^2}{{\widehat \Pi} {\widehat F}}(dt+a_k\mu_k^2
d\varphi_k)^2+\frac{{\widehat \Pi} {\widehat F}}{{\widehat \Pi}-\hat\mu
r^2}dr^2\nonumber\\
&&\qquad+ d\sigma_j^2+\sigma_j^2 d\varphi_j^2\,. \eeqa
The first line is a rotating black hole metric as in \reef{odd},
but in $d-2n$ dimensions. The remaining $2n$ dimensions, in the
second line, form a flat ${\bf R}^{2n}$ with coordinates
$(\sigma_j,\varphi_j)$. So the limiting metric is a rotating black
$2n$-brane, where the rotation is on the spherical $S^{d-2(n+1)}$
sections of the horizon.

For even $d$ the general solution is
\beq ds^2=-dt^2+r^2
d\alpha^2+(r^2+a_i^2)(d\mu_i^2+\mu_i^2 d\varphi_i^2) +\frac{\mu
r}{\Pi F}(dt+a_i\mu_i^2 d\varphi_i)^2+\frac{\Pi F}{\Pi-\mu
r}dr^2\,, \labell{even}\eeq
where now $i$ runs to $d/2$, also in
$F$ and $\Pi$ in \reef{fpi}, and $\mu_i^2+\alpha^2=1$. In this
case, the existence of a horizon is guaranteed if any one of the
spins vanishes. Taking as before $n$ spins to infinity, we find
again a rotating black $2n$-brane.

The presence of horizons for generic parameters in \reef{odd} and
\reef{even} is difficult to ascertain. If all spin parameters are
non-zero, then an upper extremality bound on a combination of the
spins arises \cite{MP}. If it is exceeded, naked singularities
appear, as in the $d=4$ Kerr black hole. One may worry that, if
all spins are turned on, and we send some of them to infinity,
then the extremal bound will always be exceeded. This is not the
case, since at the same time we are taking $\mu\to\infty$ in an
appropriate manner. In the case of even $d$, sending all
parameters but one to infinity results in a Kerr $(d-4)$-brane,
and the bound on the spin of the latter is inherited from the
higher-dimensional one. Conversely, the equation for the Kerr
horizon provides a good approximation to determine the horizon for
finite but large values of the spins $a_{j}$ (\ie all $a_j$ are
large and comparable, and $a_j\gg a_1$):
\beq r_+^2+a_1^2 -\frac{\mu}{\prod_{j>1} a_j^2}\; r_+=0\,.
\labell{approxrp}\eeq
Note that this matches the equation determining the horizon for
the full solution with the approximation $r_+\ll a_j$. Either
approach yields the bound that the minimum value for the mass
parameter which yields a regular horizon is given by
\beq \mu \gsim 2|a_1|\prod_{j>1} a_j^2\,. \labell{approxmu}\eeq

In odd $d$, one can take, at most, all but {\it two} spins to
infinity in order to get a black brane (\ie there are no black
$(d-3)$-branes in $d$ dimensions, since there are no
asymptotically flat vacuum black holes in three dimensions). In
this case the generic limiting black brane is the product of ${\bf
R}^{d-5}$ times the doubly-spinning five-dimensional black hole.
If both spins are non-zero, then there is a regular extremal
limit, while if one of them is zero the extremal solution is
singular \cite{MP}.\footnote{It is quite likely that more general
higher-dimensional solutions exist which have as a limit the
product of the black ring of \cite{ER} times a flat ${\bf
R}^{2n}$.} In each case one can obtain the analogue of
Eq.~\reef{approxrp}. In general, the condition for the existence
of an ultra-spinning regime with $n$ fast spins becomes, in the
limit, the same as the condition for the existence of a horizon in
$d-2n$ dimensions, with only slow spins, and with mass parameter
given by \reef{hatmu}.

If all the non-infinite rotations $a_k$ are zero, then the
analysis of \cite{GL} shows that the limiting static black branes
are unstable. The results of \cite{GL} have not been extended to
rotating black branes, not even to the Kerr black branes, and it
is not known the range of spins for which an instability occurs.
It seems reasonable to expect that for small rotation $a_k$ the
instability should be present, but in general a more complicated
behavior can be expected, on the basis of a conjectured connection
between classical and thermodynamic stability of black branes
\cite{GM,GWK}. This is currently under investigation.

\section{Death by radiation}
\labell{radiation}

A conservative conjecture as to the evolution of the
ultra-spinning black holes is that 
with the onset of the instability, the black hole horizon
will generically be distorted in such a way that its axial symmetry
is broken. Since
the black hole is rotating, a varying quadrupole moment will
appear and gravitational waves will be emitted. In this way, the
black hole will shed a fraction of both its spin and its mass. If
it loses enough angular momentum, it may return to a range of
values of $(M,J)$ within the stable regime. Again, this
process must proceed in a way that the horizon area increases
monotonically throughout so as to be compatible with Hawking's
area theorem.

To produce a precise description of the spin-down above, it
seems one can not avoid solving the higher-dimensional Einstein
equations in a strong field regime. Hence we were not able to
produce a detailed calculation of the gravitational radiation.
However, we present some calculations which make some
tentative steps in this direction. In particular, we have in mind
processes in which the black hole is only slightly perturbed.
For example, if the original
angular momentum is only slightly beyond the region of stability,
one might imagine that throughout the entire evolution the black
hole will only deviate from the family of solutions in eq.~\reef{mphole}.
Now as commented above, the area theorem requires that the emission
of gravitational radiation must be accompanied by an increase of
the horizon area. Having supposed that there are processes whose
effects are perturbative, we can examine the variation of the horizon
area of the solutions
\reef{mphole}. Such an analysis might give a hint as to the
boundary between the stable and unstable regimes.  However,
we warn the reader that our results are inconclusive.

We begin with some general observations. Imagine that
one detects the radiation at a large sphere surrounding the black
hole. In some short interval, it carries away some angular
momentum $\delta J$ from the black hole, and also some energy
$\delta E$. On general grounds, one might imagine that $\delta J$
is directly proportional to $\delta E$. However, on
dimensional grounds, we must introduce some physically relevant
scale in the proportionality constant. In a standard radiation
calculation, this might be the frequency of the emitted waves. In
the present case, it turns out to be convenient to use the
angular frequency of the horizon,
\beq \Omega= {a\over r_+^2+a^2}\,,
\labell{freq}\eeq
with which we write
\beq \Omega\,\delta J={\alpha}\,\delta E \,. \labell{effect} \eeq
Here $\alpha$ is a dimensionless coefficient that characterizes
the efficiency of the radiation process, \ie large $\alpha$
corresponds to a process that radiates spin efficiently.
Now for a slightly perturbed solution, the first law of black hole
mechanics allows us to obtain
the variation in the black hole area as
\beqa
\frac{\kappa}{8\pi}\,\delta \cA&=&\Omega\,\delta J-\delta E\nonumber\\
&=& (\alpha-1)\,\delta
E\,,
\eeqa
where $\kappa$ is the surface gravity of the black hole horizon.
Hence with the definition of the efficiency factor in eq.~\reef{effect}, 
we have the simple result that demanding that $\delta\cA\ge 0$ requires
\beq \alpha\ge1\,.\labell{effic}
\eeq
This result only predicts that the area theorem can be
satisfied for any spin provided that there exist radiation
processes with $\alpha>1$. Of course, we expect that this 
efficiency can only be attained for large spins, \ie
an instability is only expected to appear in the ultra-spinning
regime. To determine what realistic values for efficiency factor might
be, we examined two models for the emission of gravitational waves. 

The first was to consider the emission from a near-Newtonian slow-motion
source. In four dimensions, this is a textbook calculation \cite{MTW}
and has been recently been extended to higher dimensions in ref.~\cite{lemos}.
We present no details here but rather observe that the final result $\delta E
=\omega\,\delta J$ matches the variation in energy and angular momentum
loss of the source which rigidly rotates with angular frequency $\omega$
(in any spacetime dimension). If we assume that this result were to apply
in the present problem with $\omega=\Omega$, it would give $\alpha=1$
independent of the value of the spin parameter $a$. The latter is, of course,
unexpected since we are looking to find an instability only in the
ultra-spinning regime. It is likely that it is simply inappropriate
to consider the black hole to be a rigidly rotating body, \ie, both the angular
frequency and the moment of inertia vary during the spin-down.

We also consider a model where the rotating black hole emits
a null particle in the plane of rotation. One might think of this
process as approximating ``radiation'' focussed in the equatorial
plane. In particular, one can consider the emission of many or several
massless particles and the analysis would be essentially unchanged
with the contributions of the individual particles combining
additively. Unfortunately our final result is that this
conjectured decay mechanism does {\it not} lead to an
increase in the area of the horizons, and therefore is ruled out
as a possible decay channel. Nevertheless, it provides an
instructive example to consider, and it is also an instance where
the calculations can be performed quite rigorously, in particular
we can determine precisely the efficiency factor $\alpha$, or
equivalently, the impact parameter $R$ of the outgoing particle.
Hence we present the calculation in the following subsection.

The latter negative result is somewhat disturbing as since the null
particles are confined to the plane of rotation, this would seem
the most efficient possible way to dissipate the angular momentum.
It is hard to believe that gravitational 
radiation from distortions in an ultra-spinning horizon
would not cause a spin-down of the black hole. We take
our present results as an indication that the instability must produce
distortions deep in the nonlinear regime for the radiation process to
be efficient. That is, we can not rely on calculations which consider 
only small perturbations
of the stationary black hole solutions \reef{mphole}.

\subsection{Emission of massless particles}
\labell{zeroemission}

Consider that the rotating black hole emits a massless particle from its
edge in the plane of rotation, tangentially to the horizon and in the
direction of rotation. The particle carries away energy ${\delta E}$
and orbital angular momentum $\delta J$, small enough so we can treat it as a
test particle in the black hole background. We also consider that the
black hole evolves from the solution with mass and spin ($M$, $J$) to an
infinitesimally close one with ($M-{\delta E}$, $J-\delta J$). The trajectory
of the massless particle can be easily computed as a null geodesic in
the plane of rotation of the black
hole. For a massless particle that at infinity has energy ${\delta E}$
and orbital angular momentum $\delta J$, the radial equation is
\beq
\dot r^2=({\delta E})^2 - V(r)
\eeq
with
\beq
V(r)=-\frac{\mu \left(\delta J-a \,{\delta E}\right)^2}{r^{d-1}}+\frac{{(\delta
J)}^2-a^2 \,{(\delta E)}^2}{r^2}\,,
\eeq
and where $\mu$ and $a$ give $M$ and $J$ as in \reef{mandj}. This potential
has a maximum at a critical value
\beq
r_c=\left(
\frac{(d-1)\mu}{2}\frac{\delta J-a\,{\delta E}}{\delta J+a\,{\delta
E}}\right)^{1/(d-3)}\,.
\eeq
A particle going out from $r<r_c$ needs to have enough energy to climb up
the gravitational potential in order to escape out to infinity. This
determines a minimum energy ${(\delta E)}^2=V(r_c)$ and therefore a
maximum possible impact parameter $R=\delta J/{\delta E}$ for a particle
escaping the black hole tangentially on the plane of rotation. Then the
maximal impact parameter is determined by solving
\beq
\left(\frac{R}{a}+ 1\right)^{d-1}
\left(\frac{R}{a}-1\right)^{d-5}=\frac{1}{4}\frac{(d-1)^{d-1}}{(d-3)^{d-3}}\;
\left(\frac{\mu}{a^{d-3}}\right)^2\,.
\labell{impact}\eeq

Now comparing to the previous discussion, we see that the efficiency
factor is given by $\alpha=\Omega\,R$. Hence we must determine if this
expression can be larger than one.

In the
ultra-spinning regime the r.h.s.~of eq.~\reef{impact} is a small
quantity, and so we must have $R/a\sim 1$.\footnote{Here we assume $R$
is positive. $R<0$ and $R\sim -a$ corresponds to the particle coming in
on the other side of the center axis.} This result agrees with our previous
estimate that the geometric size of the pancaked horizon is of order
$a$ in the plane of rotation. Solving Eq.~\reef{impact} approximately 
with large $a$ yields 
\beq \alpha= {a\,R\over r_+^2+ a^2}=1+c_d\frac{r_+^2}{a^2}+O((r_+/a)^{3}) 
\qquad{\rm
where}\quad c_d=\left[\frac{1}{2^{d+1}}\frac{(d-1)^{d-1}}{(d-3)^{d-
3}}\right]^{\frac{1}{d-5}}-1\,. \labell{Rlargea}\eeq

Unfortunately, for all dimensions $d\geq 6$, we have $c_d<0$ and so, even
in the extreme ultra-spinning regime, the inequality \reef{effic}
is {\it not} satisfied. We have also examined the solutions to
eq.~\reef{impact} numerically for a wide range of $d$, and in all
cases we have found that $\alpha<1$, \ie the area never
increases by emission of the massless particle.\footnote{Closer
examination shows that this does not happen in $d=5$ either.}
Therefore, this does not seem to be a possible decay channel for
ultra-spinning black holes.

Note that the requirement $\alpha>1$, \ie $R/a>1/(a\Omega)$, is more
stringent than that in eq.~\reef{master} which, for large enough
$a/r_+$, only requires $R/a>2/(d-2)$. This can be attributed to the fact
that in the present case one of the decay products (the massless
particle) does not contribute to the total final horizon area, in
contrast to the fragmentation scenario.

It is interesting to consider this result in a different light by
reversing the arrow of time. Doing so, we would be studying the
variation in the final area obtained by throwing a massless particle at
the rotating black hole, at the maximum possible impact parameter for
capture. The analysis above shows now that if the particle is captured
by the black hole, then there is an infinitesimally close black hole of
mass $M+{\delta E}$ and spin $J+\delta J$, and with {\it larger} area, that the
system can evolve into. If there had not been any, it would have been a
clear sign of an instability -- the area theorem forbids a
decrease in the area, so a violent back reaction should have been necessary
in order for the black hole to evolve into a configuration, of larger
area, not infinitesimally close to the initial one.


\end{document}